\documentclass[twocolumn,preprintnumbers,amsmath,amssymb]{revtex4}

\usepackage{graphicx}
\usepackage{dcolumn}
\usepackage{bm}

\begin{document}

\title{Quasi one-dimensional transport in single GaAs/AlGaAs core-shell nanowires}

\author{D. Lucot}
\author{F. Jabeen}
\author{J.-C. Harmand}
\author{G. Patriarche}
\author{R. Giraud}
\author{G. Faini}
\author{D. Mailly}
\affiliation{Laboratoire de Photonique et de Nanostructures (LPN-CNRS), route de Nozay, 91460 Marcoussis, France}

\date{\today}

\begin{abstract}

We present an original approach to fabricate single GaAs/AlGaAs core-shell nanowire with robust and reproducible transport properties. The core-shell structure is buried in an insulating GaAs overlayer and connected as grown in a two probe set-up using the highly doped growth substrate and a top diffused contact. The measured conductance shows a non-ohmic behavior with temperature and voltage-bias dependences following power laws, as expected for a quasi-1D system.

\end{abstract}



\maketitle

Semiconductor nanowires (NWs) are promising candidates for the realization of innovative nano-devices for electronics as well as for photonics\cite{Li}. They also represent a new test bed system in semiconductor material science to explore the fundamental properties of 1D systems\cite{Doh}. Usual fabrication of such nano-objects by lithography and etching techniques is limited by surface damage and roughness which, at very small sizes, have a dominant effect on their physical properties. In contrast, the metal particle-mediated vapor-liquid-solid (VLS) growth mechanism allows to obtain NWs of high crystalline quality with uniform nanometer-scale diameters. Moreover, this bottom-up approach gives the possibility to achieve heterostructure material combinations that are not possible in bulk semiconductors\cite{Lauhon}. In particular core-shell heterostructures formed by the growth of crystalline overlayers around the initial NW reduces surface states which can act as scattering or recombination centers\cite{Lu}. 

The GaAs/AlGaAs material system, which presents very large band offsets and small lattice mismatches, has been studied and used extensively to fabricate heterostructures with complex band engineering along the growth direction of the crystal\cite{Capasso}. The modulation-doping concept\cite{Dingle} was first implemented in this system and high electron mobility transistors were demonstrated\cite{Mimura}. Nowadays, a low-temperature electron mobility of several $10^{6} cm^{2}/V.s$ is currently obtained in 2D structures \cite{Umansky}. To achieve 1D carrier confinement it is thus of particular interest to use the well-known GaAs/AlGaAs system and realize core-shell NW by wrapping a GaAs NW core with an AlGaAs shell layer. VLS technique has been used to produce such GaAs/AlGaAs core-shell NWs\cite{Noborisaka,Tambe,Joyce}, but up to now these heterostructures have been mainly characterized by photoluminescence\cite{Joyce, Tomioka} due to difficulty in achieving good electrical contacts. 

The standard technique to contact a NW is to mechanically detach them from the growth substrate, and after dispersion on an insulating substrate, to evaporate contact pads by lithography. This technique may suffer from the random positioning of the wires, from the oxidation of the AlGaAs shell and from the presence of pinned charges in the insulating subtrate. In this letter, we present an original method using molecular beam epitaxy (MBE) and electron-beam lithography (EBL) to achieve modulation-doped GaAs/AlGaAs core-shell NWs embedded in a GaAs matrix. The NWs are individually connected directly on their growth substrate and show a small contact resistance enabling low-temperature transport measurements. Clear signatures of their quasi-1D character are observed, opening up new opportunities for the investigation of quantum transport.

\begin{figure}[hbtp]
\includegraphics[width=0.45\textwidth]{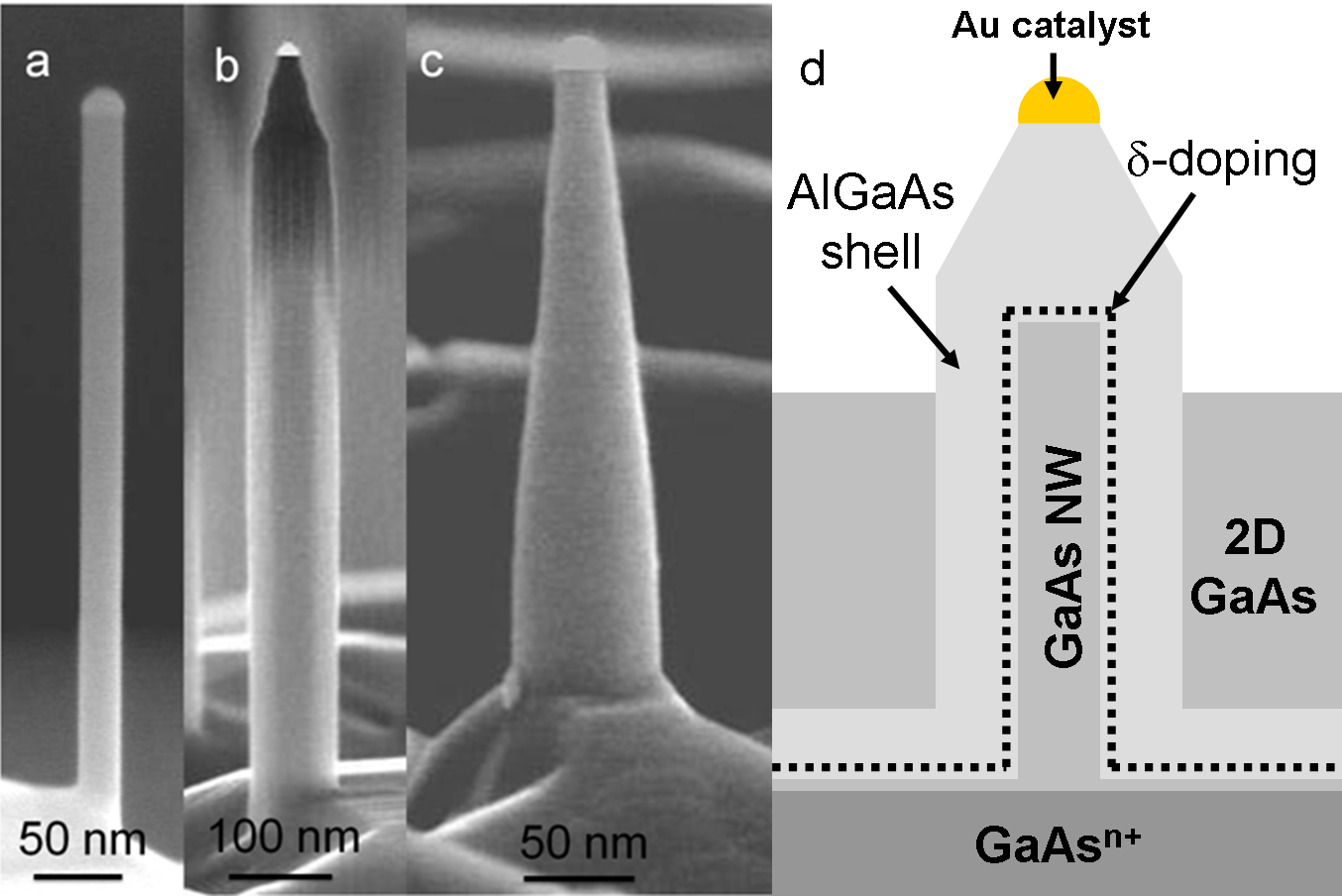}
\vskip -0.2cm
\footnotesize{\caption{Scanning Electronic Microscope (SEM) pictures of free standing 25nm diameter GaAs NWs before (a) and after (b) 40nm AlGaAs shell growth. (c) Tilted SEM micrograph of the top end of GaAs/AlGaAs core-shell NW after a partial burying in GaAs. (d) Schematics of a GaAs/AlGaAs modulation doped core-shell NW buried into a semi-insulator GaAs matrix. }}
\label{fig1}
\end{figure}

GaAs (111) B substrates are patterned with nano-sized gold dots realized by EBL and lift-off, at well defined surface locations. The samples are introduced in the MBE equipment and deoxidized at relatively low temperature ($350^{\circ}$C) under an atomic hydrogen flux. Then, the sample temperature is increased up to $550^{\circ}$C for Au-assisted VLS growth\cite{Harmand}. At this temperature, the gold dots are alloyed with Ga and form small droplets. The exact size of the droplet is determined by the original diameter of the EBL patterning, the thickness of the Au film and the growth conditions. Ga and As$_{4}$ fluxes are supplied to form nominally undoped vertical GaAs NWs by VLS, in epitaxial relationship with the substrate. At this stage, the crystalline structure of these NWs is wurtzite-like with several stacking faults\cite{Patriarche}. The NWs present a regular hexagonal cross-section limited by (1010) facets. Along their length, they have a constant lateral size, close to the diameter of the catalyst drop (Fig. 1a). This diameter can be easily tuned between 20 nm and 100 nm and the NW length, fixed by the growth time, is about $1 \mu m$. Then, an AlGaAs shell is formed on the NW facets (Fig. 1b). The sidewall growth is promoted by the presence of Al and by swapping the As$_{4}$ flux to As$_{2}$ flux, the adatom diffusion length on the sidewall facets being strongly reduced under these conditions\cite{Sartel}. The nominal Al composition of the shell is $30 \%$. A first radial layer of 10 nm is formed, then Si flux is supplied to give a Si $\delta$-doping with a nominal sheet concentration of $1\times10^{12}cm^{-2}$ on the facets. The shell is completed with a 30 nm AlGaAs layer. Hence, in this structure, we expect a transfer of free carriers in the GaAs core where no doping impurities are introduced. Because it results from epitaxial nucleation on the NW sidewalls, the shell adopts the wurtzite structure of the core. It has to be noted that during the shell formation, similar layers deposit on the substrate surface and on top of the NWs (Fig. 1d). 

Finally, the resulting core-shell NWs are buried (Fig. 1c) by an undoped GaAs overgrowth performed at higher temperature ($630^{\circ}$C). At this temperature, the VLS growth is inhibited and there is no lateral growth on the NW sidewalls. GaAs grows only on the substrate surface, adopting its zinc blende bulk structure and embedding progressively the NWs. We have shown that this process transforms the initial wurtzite crystal phase of the NWs in the zinc blende phase adopted by the burying layer\cite{Patriarche}. The phase transformation is very favorable in that it eliminates the stacking faults pre-existing in the NWs. We thus end up with perfect cubic core-shell NWs embedded in a cubic matrix. Since they no longer have free sidewall facets, these NWs are perfectly protected against aging by surface oxidation and their electronic properties are not influenced by surface states. This is confirmed by the excellent reproducibility of our transport measurements over several months. 

\begin{figure}[hbtp]
\includegraphics[width=0.5\textwidth]{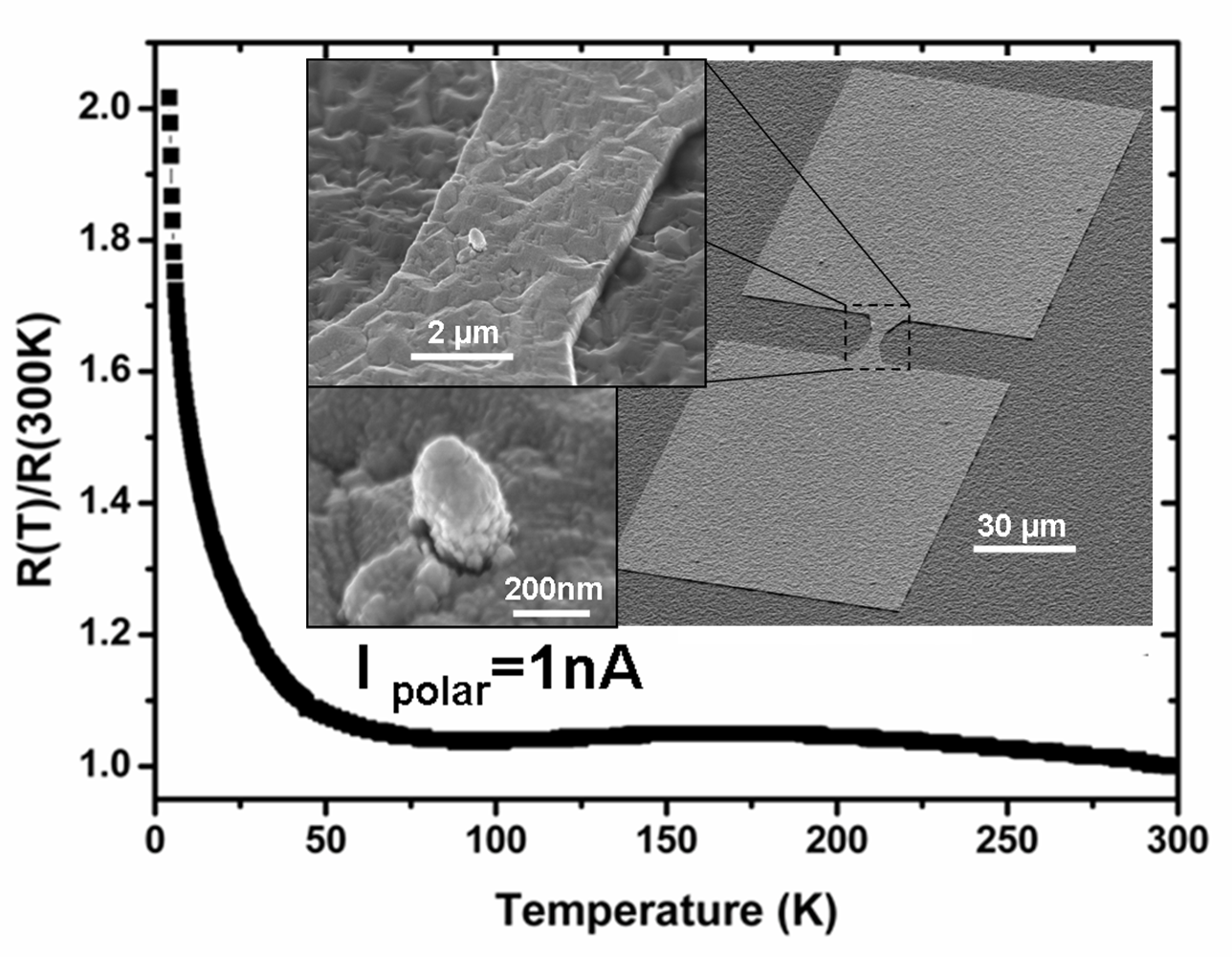}
\vskip -0.1cm
\footnotesize{\caption{Typical temperature dependence of the resistance R (T) normalized to R (300 K) for a single GaAs/AlGaAs core-shell NW with a 25 nm GaAs core diameter. (inset) SEM picture of a single NW partially buried in an undoped GaAs matrix and electrically contacted to a Ni/Ge/Au electrode at its top end.}}
\label{fig2} 
\end{figure}

The planarization of the sample induced by the GaAs layer overgrowth enables the connection of single NWs directly on their substrate (Fig. 1d). A first contact is obtained on the emerging part of the NW by using a conventional Ni/Ge/Au film deposition patterned by EBL (Fig. 2. inset). A second one is directly taken on the backside of the highly doped n-type substrate. After a rapid (5 s) thermal annealing of the device at $430^{\circ}$C, we obtained  good ohmic current-voltage (I-V) characteristics with a total two-probe resistance ranging from $400\Omega$ to $2k\Omega$ at room temperature (RT), for all the 12 different single NWs of 25 nm GaAs core diameter investigated\cite{comment}.    

The electrical properties of single NWs are measured in this two-probe configuration in a $^{4}He$ VTI cryostat using current-biased lock-in techniques. The first important observation is that the resistance of the samples is significantly lower than the quantum of resistance ($h/2e^{2}\sim12.9k\Omega$) expected for a single propagating mode connected to ideal contacts. This suggests that we obtain high-transparency contacts and that a single NW sustains several propagating modes, similarly to the case of multi walled carbon nanotube (CNT) \cite{Liu}. A good contact coupling to NW is of particular interest because it sets our device in a conduction regime where classical Coulomb charging effects due to the presence of large tunnel barriers are negligible \cite{Liu,Bockrath}. Moreover, the distinction between two- and four-terminal resistance is less pertinent and the two-points configuration used in the present study is expected to capture most of the intrinsic physics. 
 
Now we focus on the temperature dependence of the NW resistance. As shown on figure 2, two conduction regimes can be distinguished. As the temperature is decreased from RT to 80 K, the resistance exhibits little variations. A similar behavior has been already reported in Si/Ge core-shell NWs \cite{Lu} and is ascribed to the weak acoustic phonon contribution in 1D systems. Below 80 K, the resistance increases while lowering the temperature. This increase cannot be explained by a strong localization effect. Indeed, we cannot fit our experimental data with an activation law $(ln(R)\propto1/T)$ or a variable range hopping behaviour $(ln(R)\propto1/T^{1/2}$, for a 1D wire), confirming the absence of any large tunnel barrier within or at the ends of the NW.

\begin{figure}[hbtp]
\includegraphics[width=0.4\textwidth]{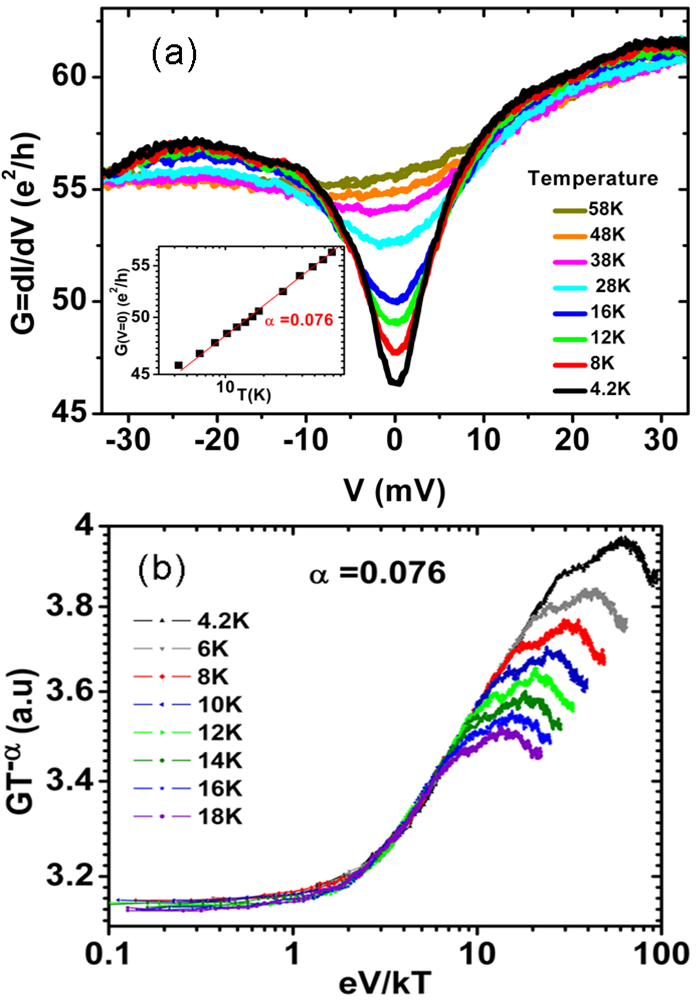}
\vskip -0.2cm
\footnotesize{\caption{(a) Differential conductance G = dI/dV of a single NW as a function of bias voltage for different temperatures. The inset shows a log-log representation of G(V=0) versus T. The dependence follows a power-law with the exponent $\alpha=0.076$. (b) $GT^{-\alpha}$ versus $eV/k_{B}T$ plot of the same data as in (a), using $\alpha=0.076$.}}
\label{fig3}
\end{figure}

The low temperature behavior is further investigated by measuring the differential conductance $G=dI/dV$ of a single NW device as function of the bias-voltage, for temperature ranging from 65 to 4.2 K (Fig. 3a). A conductance dip centered at zero bias (zero-bias anomaly) appears and grows up as the temperature is decreased. As shown in the Fig. 3a insert, the temperature dependence of the zero-bias conductance ($G_{V=0}$) is well described by a power law $G_{V=0}\propto T^{\alpha}$ with $\alpha = 0.076$. In the same way the voltage dependence can also be described by a power law $G_{V}\propto V^{\alpha}$ with the same exponent $\alpha = 0.076$, for bias voltages larger than $k_{B}T/e$. Both these laws can be resumed by a scaling law $GT^{-\alpha}=f(eV/k_{B}T)$ where $f$ is the scaling function and $\alpha$ is a scaling coefficient specific to the measured NW. Thus, the G(V) curves measured at different temperatures collapse remarkably well on to a single curve $GT^{-\alpha}$ vs $eV/k_{B}T$, as depicted in Fig. 3b. All the measured NWs exhibit a scaling law with an $\alpha$ coefficient distributed from 0.02 to 0.23 and with no clear correlation to their RT resistance.

Such scaling laws have been previously observed on many other low dimensional systems including single or multiwalled CNTs \cite{Liu,Bockrath, Bachtold}, metallic quantum wire \cite{Slot,Venkataraman} or split-gated NWs formed on planar GaAs/AlGaAs heterostructures \cite{Jompol}. This behaviour is related to the energy cost for an electron from the 3D reservoir to tunnel into a 1D system\cite{Deshpande}. Indeed, adding an electron to a pure 1D conductor requires changing the many body state of its collective excitations. It yields to a vanishing electron tunnelling density of states at low energy leading to a power law dependence of the conductance on V and T. The exponent $\alpha$ depends principally on the electron-electron interaction strength and on the number of conducting channels \cite{Venkataraman}. For a clean single 1D propagating channel, electron-electron interaction is expected to give $\alpha\approx0.5$\cite{Jompol}. The lower $\alpha$ values extracted here are similar to those obtained for multi-wall CNTs \cite{Liu,Bachtold} and indicate that our NWs are quasi-1D systems with a large number of 1D conducting channels in parallel.

In conclusion, we have developed an original fabrication method to obtain buried GaAs/AlGaAs core-shell NWs with a Si $\delta$-doping in the AlGaAs shell showing a clear quasi-1D behavior. Their geometrical, structural and electrical characteristics are easily tunable (as compared to the case of CNTs for instance) by adjusting the process parameters. The transport properties do not evolve with time and the direct contacting of the NW on the growth substrate allows one to create a complex routing of connections leading to large scale integration. Thus, these new classes of materials are of broad interest for future applications in nanoelectronics, and are well-suited to study phase coherent quantum transport in one-dimension.

The authors thank Jean-Eric Wegrowe, Ulf Gennser, for helpful discussions and Laurent Travers, Christian Ulysse for their technical assistance. This work was partly supported by the R\'egion Ile-de-France in the framework of C'nano IdF.


\begin{references}
\bibitem{comment} We checked the insulating behavior of the burying layer by measuring the resistance between the bottom contact and a top contact deposited on a sample surface area with no emerging NW. In that case, the resistance is always higher than $10$M$\Omega$.
\end{references}

\end{document}